\documentclass[12pt]{article}
\usepackage{graphicx}

      \usepackage[cp1251]{inputenc} 

\begin{document}
 \noindent {\footnotesize\it Astronomy Letters, 2013, Vol. 39, No. 12, pp. 819--825.}
 \newcommand{\dif}{\textrm{d}}

 \noindent
 \begin{tabular}{llllllllllllllllllllllllllllllllllllllllllllll}
 & & & & & & & & & & & & & & & & & & & & & & & & & & & & & & & & & & & & & \\\hline\hline
 \end{tabular}

 \vskip 1.0cm

 \centerline{\bf Cepheid Kinematics and the Galactic Warp}
 \bigskip
 \centerline{\bf V.V. Bobylev}
 \bigskip
 \centerline{\small \it Pulkovo Astronomical Observatory, St. Petersburg,  Russia}
 \centerline{\small \it Sobolev Astronomical Institute, St. Petersburg State University, Russia}
 \bigskip
 \bigskip
{\bf Abstract}—The space velocities of 200 long-period
($P>5$~days) classical Cepheids with known proper motions and
line-of-sight velocities whose distances were estimated from the
period--luminosity relation have been analyzed. The linear
Ogorodnikov-Milne model has been applied, with the Galactic
rotation having been excluded from the observed velocities in
advance. Two significant gradients have been found in the Cepheid
velocities,
 $\partial W/\partial Y=-2.1\pm0.7$~km s$^{-1}$ kpc$^{-1}$ and
 $\partial V/\partial Z=  27\pm10 $~km s$^{-1}$ kpc$^{-1}$.
 In such a case, the angular velocity of solid-body
rotation around the Galactic $X$ axis directed to the Galactic
center is $-15\pm5$~km s$^{-1}$ kpc$^{-1}$.


\section*{INTRODUCTION}
As analysis of the large-scale structure of neutral hydrogen
showed, a warp of the gas disk is observed in the Galaxy
(Westerhout 1957). The results of studying this structure using
the currently available data on the HI and HII distributions are
presented in Kalberla and Dedes (2008) and Cersosimo et al.
(2009), respectively. The warp is seen in the distribution of
stars and dust (Drimmel and Spergel 2001), pulsars (Yusifov 2004),
OB stars from the Hipparcos catalogue (Miyamoto and Zhu 1998), and
in the distribution of 2MASS red-giant-clump stars (Momany et al.
2006). The system of Cepheids also exhibits a similar feature
(Fernie 1968; Berdnikov 1987; Bobylev 2013).

Of great interest are the attempts to find a relationship between
the kinematics of stars and the disk warp (Miyamoto et al. 1993;
Miyamoto and Zhu 1998; Drimmel et al. 2000; Bobylev 2010). In
particular, based on the proper motions of O--B5 stars, Miyamoto
and Zhu (1998) found a positive rotation of this system of stars
around the Galactic x axis with an angular velocity of about
$+$4~km s$^{-1}$ kpc$^{-1}$. In contrast, based on the proper
motions of about 80 000 red-giant-clump stars, Bobylev (2010)
found an opposite rotation of this system of stars around the $x$
axis with an angular velocity of about $-4$~km s$^{-1}$
kpc$^{-1}$.

The stellar proper motions alone do not allow complete information
to be obtained. In this respect, although the Cepheids are not all
that many, they are a unique tool for studying the
three-dimensional kinematics of the Galaxy: the distances, proper
motions, and line-of-sight velocities are known for them.

A number of models were proposed to explain the Galactic warp: (1)
the interaction between the disk and a nonspherical dark matter
halo (Sparke and Casertano 1988); (2) the gravitational influence
from the Galaxy's nearest satellites (Bailin 2003); (3) the
interaction of the disk with the flow near the Galaxy formed by
high-velocity hydrogen clouds that resulted from mass exchange
between the Galaxy and the Magellanic Clouds (Olano 2004); (4) the
intergalactic flow (L\'opez-Corredoira et al. 2002); and (5) the
interaction with the intergalactic magnetic field (Battaner et al.
1990).

Note that the term ``warp'' implies some nonlinear dependence.
However, we attempt to find a relationship between the kinematics
of stars and the warp of the hydrogen layer in the form of a
simple linear approach. For this purpose, we search, for example,
for the rotation of the symmetry plane of the system of stars
around some axis. Since the symmetry plane of the Cepheid system
is inclined to the Galactic plane at an angle of $\approx-2^\circ$
in a direction of $\approx$270$^\circ$ (Bobylev 2013), the most
suitable manifestation of the relationship is the rotation of the
system around the Galactic $x$ axis.

The goal of this study is to reveal the relationship between the
Cepheid velocities and the warp of the stellar-gaseous Galactic
disk. For this purpose, we use a sample of long-period classical
Cepheids with measured proper motions and line-of-sight velocities
and estimate their distances from the period–luminosity relation.
We apply the linear Ogorodnikov-Milne model for our analysis and
exclude the Galactic rotation from the observed velocities in
advance, focusing our attention on the motion in the $XZ$ and $YZ$
planes.

\section*{DATA}
We use Cepheids of the Galaxy’s flat component classified as DCEP,
DCEPS, CEP(B), CEP in the GCVS (Kazarovets et al. 2009) as well as
CEPS used by other authors. To determine the distance based on
from the period–luminosity relation, we used the calibration from
Fouqu\'e et al. (2007):
 $\langle M_V\rangle=-1.275-2.678 \log P,$ where the period $P$ is in days.
Given $\langle M_V\rangle,$ taking the period-averaged apparent
magnitudes $\langle V\rangle$ and extinction
 $A_V=3.23 E(\langle B\rangle-\langle V\rangle)$ mainly from Acharova et al. (2012)
and, for several stars, from Feast and Whitelock (1997), we
determine the distance $r$ from the relation
 \begin{equation}\displaystyle
 r=10^{\displaystyle -0.2(\langle M_V\rangle-\langle V\rangle-5+A_V)}.
 \label{Ceph-02}
 \end{equation}
For a number of Cepheids (without extinction data), we used the
distances from the catalog by Berdnikov et al. (2000) determined
from infrared photometry.

Data from Mishurov et al. (1997) and Gontcharov (2006) as well as
from the SIMBAD and DDO databases served as the main sources of
line-of-sight velocities for Cepheids. As a rule, the proper
motions were taken from the UCAC4 catalog (Zacharias et al. 2013)
and, in several cases, from TRC (Hog et al. 2000).

Proceeding from the goals of our study, we concluded that it would
be better not to use several stars located above the Galactic
plane by more than 2 kpc and deep in the inner Galaxy. Thus, we
used the constraints
 \begin{equation}
  \begin{array}{c}
    |Z|<2~\hbox {kpc}, \\
      P>5^d,\\
 |V_{pec}|<100~\hbox {km s$^{-1}$},\\
  \sigma_V<80~\hbox {km s$^{-1}$},
 \label{criterii-xz}
 \end{array}
 \end{equation}
satisfied by 205 Cepheids. When calculating the velocity errors,
we assumed the distance error to be 10\%. In particular, the
constraint for $\sigma_V$ in (2) is the random error in the total
space velocity of a star. The constraint on the pulsation period P
was chosen from the following considerations. Our analysis of the
distribution of classical Cepheids (Bobylev 2013) shows that the
oldest Cepheids with periods $P<5^d$ have a significantly
different orientation than younger Cepheids.

\begin{figure}[t]
{\begin{center}
 \includegraphics[width=0.6\textwidth]{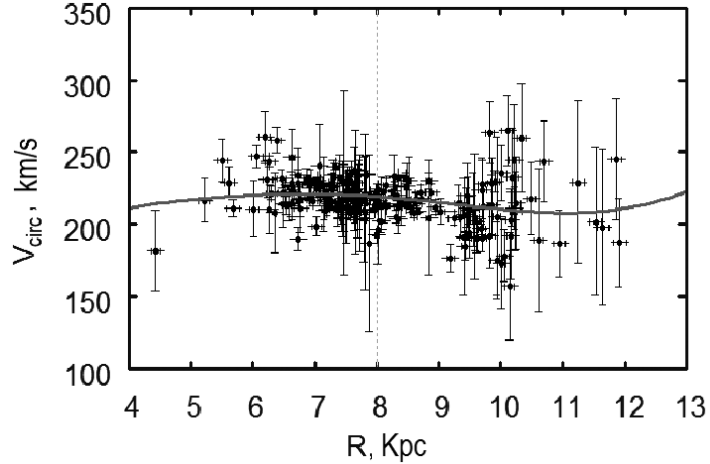}
 \caption{
Galactic rotation curve constructed with parameters~(3) (solid
line). The dotted line marks the position of the Sun. The circles
with error bars indicate the Cepheid rotation velocities.}
 \label{f1}
\end{center}}
\end{figure}

Bobylev et al. (2008) found the parameters of the Galactic
rotation curve containing six terms of the Taylor expansion of the
angular velocity of Galactic rotation $\Omega_0$ for the
Galactocentric distance of the Sun $R_0=7.5$~kpc. Data on hydrogen
clouds at tangential points, on massive star-forming regions, and
on the velocities of young open star clusters were used for this
purpose. The more up-to-date value of $R_0$ is 8~kpc (Foster and
Cooper 2010). Therefore, the parameters of the Galactic rotation
curve were redetermined using the same sample but for $R_0=8$~kpc:
 \begin{equation}
  \begin{array}{lll}
 \Omega_0  =  -27.4\pm 0.6~\hbox {km s$^{-1}$ kpc$^{-1}$},    \\
 \Omega^1_0= ~~3.80\pm0.07~\hbox {km s$^{-1}$ kpc$^{-2}$},  \\
 \Omega^2_0= -0.650\pm0.065~\hbox {km s$^{-1}$ kpc$^{-3}$}, \\
 \Omega^3_0=~~0.142\pm0.036~\hbox {km s$^{-1}$ kpc$^{-4}$}, \\
 \Omega^4_0= -0.246\pm0.034~\hbox {km s$^{-1}$ kpc$^{-5}$}, \\
 \Omega^5_0=~~0.109\pm0.020~\hbox {km s$^{-1}$ kpc$^{-6}$}.
   \label{Omega}
  \end{array}
 \end{equation}
Based on a sample of Cepheids, Bobylev and Bajkova (2012) found
 $\Omega_0 = -27.5\pm0.5$~km s$^{-1}$ kpc$^{-1}$,
 $\Omega^{'}_0 =  4.12\pm0.10$~km s$^{-1}$ kpc$^{-2}$ and
 $\Omega^{''}_0 = -0.85\pm0.07$~km s$^{-1}$ kpc$^{-3},$
 which are in good
agreement with the corresponding values (3). At the same time, the
parameters (3) allow the Galactic rotation curve to be constructed
in a wider range of Galactocentric distances $R.$ This rotation
curve is shown in Fig.~1. The parameters (3) were used to analyze
the peculiar velocity $V_{pec}$ in (2). The constraint on the
magnitude of $V_{pec}$ is an indirect constraint on the radius of
the sample, which is $r\approx6$~kpc is our case.

\section*{THE MODEL}
We use a rectangular Galactic coordinate system with its axes
directed from the observer toward the Galactic center (the $x$
axis or axis 1), in the direction of Galactic rotation (the $y$
axis or axis 2), and toward the North Galactic Pole (the $z$ axis
or axis 3).

We apply the linear Ogorodnikov–Milne model (Ogorodnikov 1965),
where the observed velocity ${\bf V}(r)$ of a star with a
heliocentric radius vector ${\bf r}$ is described, to terms of the
first order of smallness $r/R_0\ll 1,$ by the vector equation
\begin{equation}
 {\bf V}(r)={\bf V}_\odot+M{\bf r}+{\bf V'},
 \label{eq-1}
 \end{equation}
Here, ${\bf V}_\odot(X_\odot,Y_\odot,Z_\odot)$ is the Sun’s
peculiar velocity relative to the stars under consideration, {$\bf
V'$} is the star’s residual velocity, $M$ is the displacement
matrix (tensor) whose components are the partial derivatives of
the velocity ${\bf u}(u_1,u_2,u_3)$ with respect to the distance
${\bf r}(r_1, r_2, r_3),$ where ${\bf u}={\bf V}(R)-{\bf V}(R_0)$,
$R$ and $R_0$ are the Galactocentric distances of the star and the
Sun, respectively. Then,
\begin{equation}
 M_{pq}={\left(\frac{\partial u_p} {\partial r_q}\right)}_\circ, \quad p,q=1,2,3,
 \label{eq-2}
 \end{equation}
taken at $R=R_0.$ All nine elements of the matrix $M$ can be
determined using three components of the observed velocities ---
the line-of-sight velocities $V_r$ and stellar proper motions
$\mu_l\cos b,$ $\mu_b:$
 \begin{equation}
  \begin{array}{lll}
 V_r=-X_{\odot}\cos b\cos l- Y_{\odot}\cos b\sin l-Z_{\odot}\sin b +  \\
   +r[\cos^2 b\cos^2 l M_{11}+\cos^2 b\cos l\sin l M_{12}+\cos b\sin b \cos l M_{13}+\\
   +\cos^2 b\sin l\cos l M_{21}+ \cos^2 b\sin^2 l M_{22}+\cos b\sin b\sin l M_{23} +\\
   +\sin b\cos b\cos lM_{31}+\cos b\sin b\sin lM_{32}+\sin^2 b M_{33}],\\
  4.74 r \mu_l\cos b= X_\odot\sin l-Y_\odot\cos l+  \\
 +r [-\cos b\cos l\sin l  M_{11} -\cos b\sin^2 l M_{12}-\sin b \sin l M_{13}+\\
 +\cos b\cos^2 l M_{21}+\cos b\sin l\cos l M_{22}+\sin b\cos l  M_{23} ],\\
 4.74 r \mu_b=X_\odot\cos l\sin b+ Y_\odot\sin l\sin b-Z_\odot\cos b+ \\
 +r [-\sin b\cos b\cos^2 l M_{11}- \sin b\cos b\sin l \cos l M_{12}- \sin^2 b \cos l  M_{13}-\\
  -\sin b\cos b\sin l\cos l M_{21}- \sin b\cos b\sin^2 l  M_{22} -\sin^2 b\sin l  M_{23}+   \\
 +\cos^2 b\cos l M_{31} +\cos^2 b\sin l M_{32}+  \sin b\cos b  M_{33} ].
   \label{eq-5}
  \end{array}
 \end{equation}
It is useful to divide the matrix $M$ into its symmetric,
$M^{\scriptscriptstyle+}$ (local deformation tensor), and
antisymmetric, $M^{\scriptscriptstyle-}$ (rotation tensor), parts:
 \begin{equation}
 \renewcommand{\arraystretch}{2.2}
  \begin{array}{lll}\displaystyle
 M_{\scriptstyle pq}^{\scriptscriptstyle+}=
 {1\over 2}\left( \frac{\partial u_{p}}{\partial r_{q}}+
 \frac{\partial u_{q}}{\partial r_{p}}\right)_\circ,  \quad
 \displaystyle
 M_{\scriptstyle pq}^{\scriptscriptstyle-}=
 {1\over 2}\left(\frac{\partial u_{p}}{\partial r_{q}}-
 \frac{\partial u_{q}}{\partial r_{p}}\right)_\circ, \quad
  p,q=1,2,3,
 \label{eq-6}
 \end{array}
 \end{equation}
where the subscript 0 means that the derivatives are taken at
$R=R_0.$ The quantities
$M_{\scriptscriptstyle32}^{\scriptscriptstyle-},
 M_{\scriptscriptstyle13}^{\scriptscriptstyle-}$ and
 $M_{\scriptscriptstyle21}^{\scriptscriptstyle-}$ are the
components of the solid-body rotation vector of a small solar
neighborhood around the $x, y, z$ axes, respectively. In
accordance with our chosen rectangular coordinate system, the
positive rotations are those
 from axis 1 to axis 2 ($\Omega_z$),
 from axis 2 to axis 3 ($\Omega_x$), and
 from axis 3 to axis 1 ($\Omega_y$):
 \begin{equation}
 M^{\scriptscriptstyle-}= \pmatrix
  {         0&-\Omega_z & \Omega_y\cr
     \Omega_z&         0&-\Omega_x\cr
    -\Omega_y& \Omega_x&         0\cr}.
 \label{Omega-0}
 \end{equation}
The quantity $M_{\scriptscriptstyle21}^{\scriptscriptstyle-}$ is
equivalent to the Oort constant $B.$ Each of the quantities
  $M_{\scriptscriptstyle12}^{\scriptscriptstyle+},
  M_{\scriptscriptstyle13}^{\scriptscriptstyle+}$ and
  $M_{\scriptscriptstyle23}^{\scriptscriptstyle+}$ describes the deformation in the corresponding
plane; in particular,
$M_{\scriptscriptstyle12}^{\scriptscriptstyle+}$ is equivalent to
the Oort constant $A.$ The diagonal elements of the local
deformation tensor
  $M_{\scriptscriptstyle11}^{\scriptscriptstyle+},
  M_{\scriptscriptstyle22}^{\scriptscriptstyle+}$ and
  $M_{\scriptscriptstyle33}^{\scriptscriptstyle+}$
 describe the general
local compression or expansion of the entire stellar system
(divergence). The set of conditional equations (6) includes twelve
sought-for unknowns to be determined by the least-squares method.

{
\begin{table}[t]                                                
\caption[]{\small\baselineskip=1.0ex\protect
 Kinematic parameters of the Ogorodnikov–Milne model
 }
\begin{center}
\label{t1}
\begin{tabular}{|c|r|r|c|}\hline
  Parameter &  Without correction  & Wit correction  \\\hline

 $X_\odot$ & $  6.1\pm1.4$ & $  6.3\pm1.5$ \\
 $Y_\odot$ & $ 11.0\pm1.4$ & $ 10.0\pm1.5$ \\
 $Z_\odot$ & $  5.3\pm1.4$ & $  6.0\pm1.5$ \\

 $M_{11}$ & $ -0.5\pm 1.0$ & $ 0.1\pm 1.0$ \\
 $M_{12}$ & $ -2.2\pm 0.7$ & $-2.9\pm 0.7$ \\
 $M_{13}$ & $ -8.0\pm11.7$ & $ 1.9\pm11.9$ \\

 $M_{21}$ & $  1.6\pm 1.0$ & $ 0.8\pm 1.0$ \\
 $M_{22}$ & $ -0.9\pm 0.7$ & $-1.1\pm 0.7$ \\
 $M_{23}$ & $ 34.4\pm11.7$ & $32.4\pm11.9$ \\

 $M_{31}$ & $ -2.4\pm 1.0$ & $-2.8\pm 1.0$ \\
 $M_{32}$ & $ -1.1\pm 0.7$ & $-2.1\pm 0.7$ \\
 $M_{33}$ & $ 11.7\pm11.7$ & $13.2\pm11.9$ \\\hline

\end{tabular}
\end{center}
 \small\baselineskip=1.0ex\protect
Note. The velocities $X_\odot,$ $Y_\odot,$ and $Z_\odot$ are in km
s$^{-1};$ the remaining parameters are in km s$^{-1}$ kpc$^{-1}$.
\end{table}
}
\begin{figure}[t]
{\begin{center}
 \includegraphics[width=0.88\textwidth]{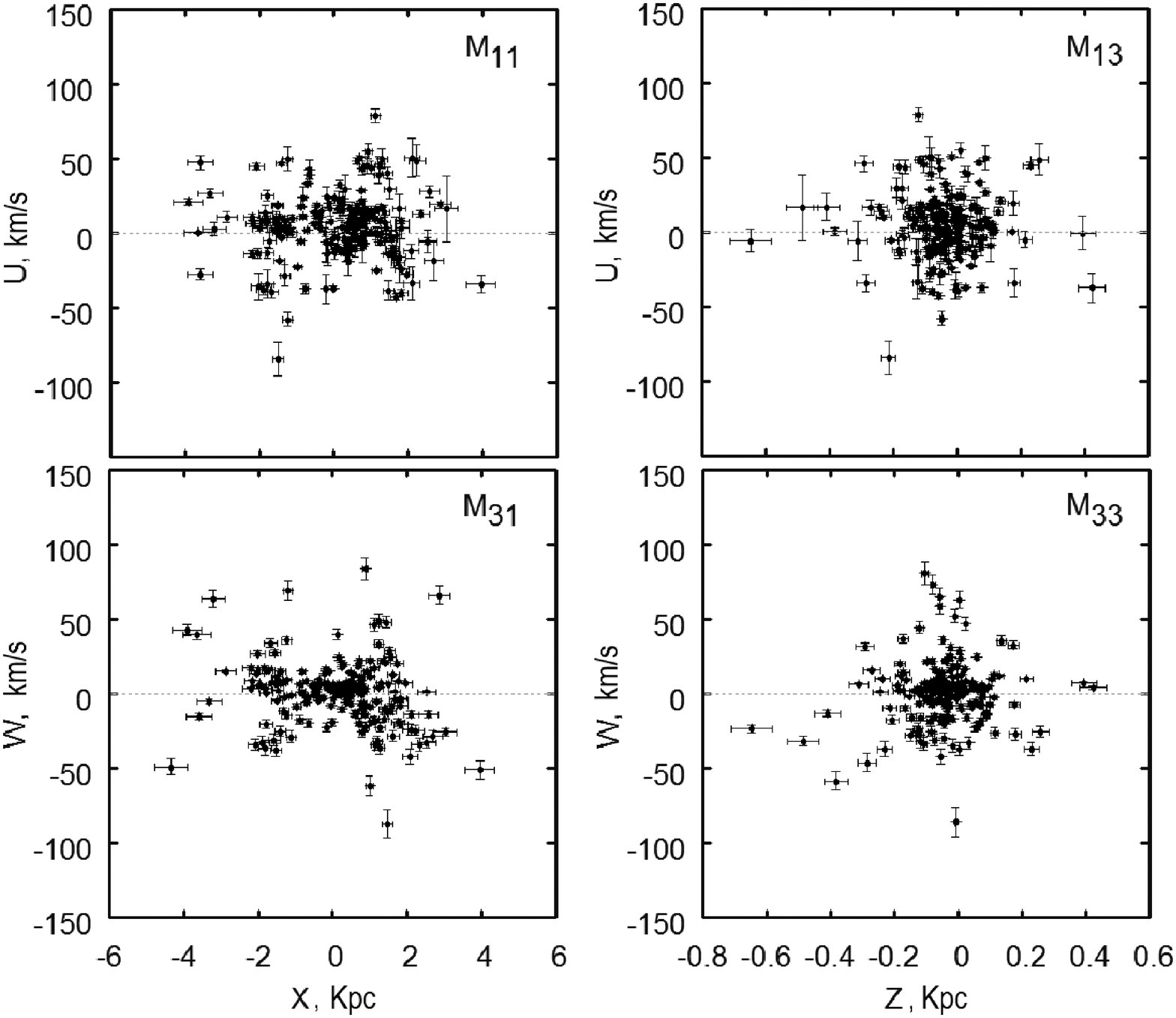}
 \caption{
Dependences describing the kinematics in the $XZ$ plane.}
 \label{f2}
\end{center}}
\end{figure}
\begin{figure}[t]
{\begin{center}
 \includegraphics[width=0.88\textwidth]{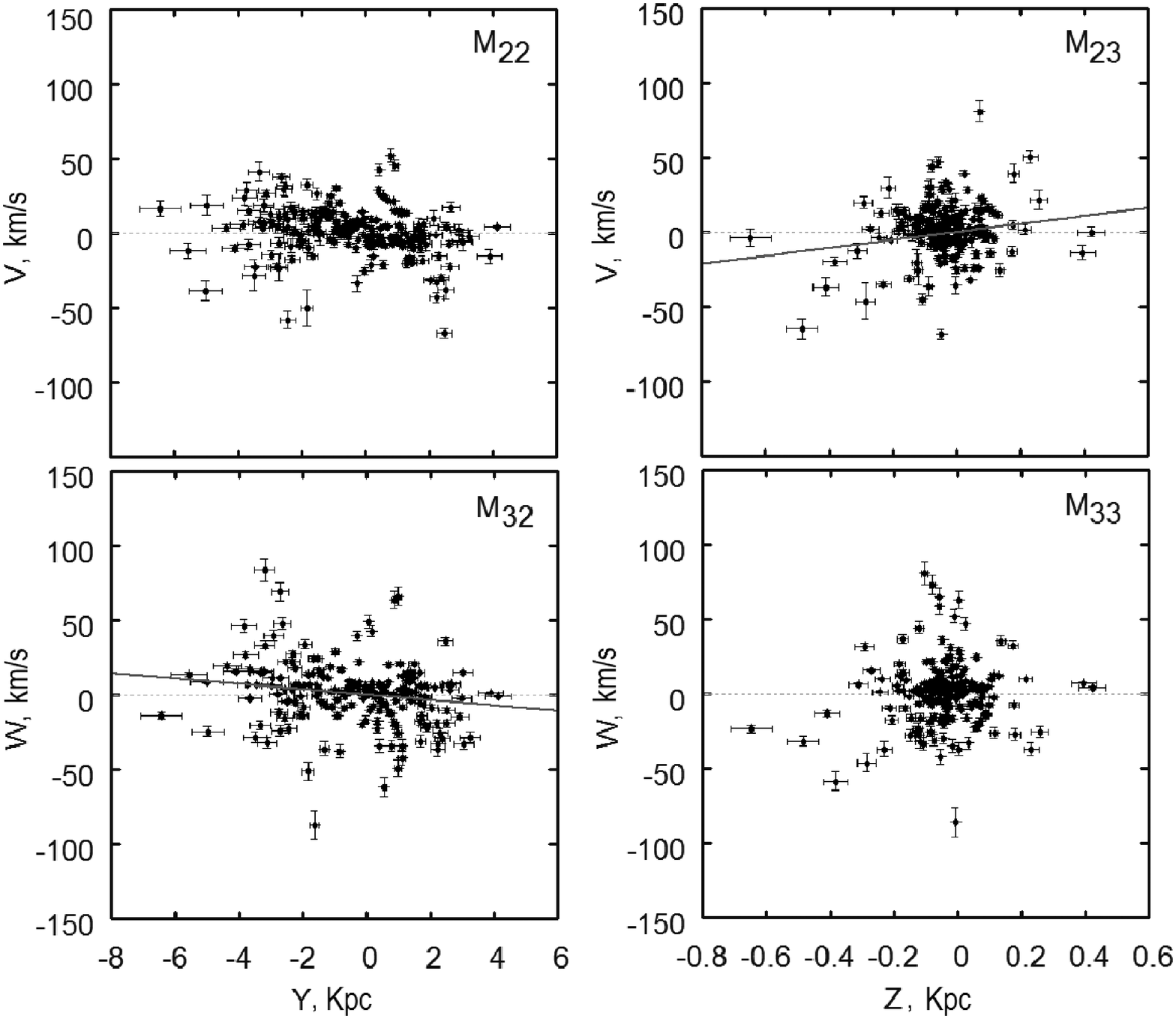}
 \caption{
Dependences describing the kinematics in the $YZ$ plane. }
 \label{f3}
\end{center}}
\end{figure}

\section*{RESULTS AND DISCUSSION}
The table gives the parameters of the Ogorognikov-Milne model
found by simultaneously solving the set of equations (6) using a
sample of 200 Cepheids. Two solutions are presented. The point is
that the ICRS/Hipparcos (1997) system, whose extension is the
UCAC4 catalog we use, has a small residual rotation relative to
the initial frame of reference. The equatorial components of this
vector are
 $(\omega_x,\omega_y,\omega_z)=(-0.11,0.24,-0.52)\pm(0.14,0.10,0.16)$ mas yr$^{-1}$
 (Bobylev 2010). Therefore, the table gives the parameters
calculated for two cases: when the Cepheid proper motions were not
corrected and when they were derived from the stellar proper
motions after applying the correction $\omega_z=-0.52$~mas
yr$^{-1}$.

There are no significant differences between the two solutions.
However, it can be noted that with the corrected proper motions,
the parameter $M_{13}$ decreased to zero and the parameter
$M_{32}$ slightly increased, which is important to us. Therefore,
below we will use the results from the last column of the table.

{\bf The XY plane.} Since $M_{12}=-\Omega_0,$ the value of
$M_{12}=-2.9\pm0.7$~km s$^{-1}$ kpc$^{-1}$ found shows that the
Cepheid velocities were slightly overcorrected (we should have
used $\Omega_0\approx-26$~km s$^{-1}$ kpc$^{-1}$ precisely for
this sample). This is of no serious importance for the goals of
our study, because this is just a linear shift, while the
nonlinear character of the Galactic rotation curve was taken into
account well. The remaining parameters describing the kinematics
in the $XY$ plane, $M_{11}, $ $M_{21}$ and $M_{22}$, are close to
zero.

{\bf The XZ plane.} As can be seen from the table, none of the
coefficients $M_{11},$ $M_{13},$ $M_{31}$ and $M_{33},$
describing the kinematics in this plane differs significantly from
zero. Figure 2 displays the corresponding distributions of stars.

{\bf The YZ plane.} Figure 3 shows the distributions of stars; the
solid lines indicate two dependences plotted according to the data
from the table:
 $M_{23}=\partial V/\partial Z=32.4\pm11.9$~km s$^{-1}$ kpc$^{-1}$ and
 $M_{32}=\partial W/\partial Y=-2.1\pm 0.7$~km s$^{-1}$ kpc$^{-1}$.
We refined the coefficient
 $M_{23}= 26.8\pm10.2$~km s$^{-1}$ kpc$^{-1}$
 using a graphical method. For this purpose, we
calculated the dependence $V=f(Z)$ from the data of the
corresponding graph in Fig.~3 with the constraint $|Z|>0.040$~kpc
(136 stars were used).

Let us now consider the displacement tensor $M_W$ that we
associate with the influence of the disk warp on the motion of the
Cepheid system:
 \begin{equation}
 M_W=  \pmatrix{
 {\strut \displaystyle\partial V}\over{\displaystyle\partial Y}& {\strut \displaystyle\partial V}\over{\displaystyle\partial Z}\cr
 {\strut \displaystyle\partial W}\over{\displaystyle\partial Y}& {\strut \displaystyle\partial W}\over{\displaystyle\partial
 Z}\cr}.
 \end{equation}
According to the data from the table, both of its diagonal
elements can be set equal to zero. This means that there are no
motions like expansion--compression in this plane. Then,
 \begin{equation}
 M_W=\pmatrix
 {           0& 26.8_{(10.2)}\cr
  -2.1_{(0.7)}& 0          \cr},
 \label{MM1}
  \end{equation}
the deformation tensor (7) takes the form
 \begin{equation}
 M_W^{\scriptscriptstyle+}= \pmatrix
  {           0&12.4_{(5.1)}\cr
   12.4_{(5.1)}&0\cr},
 \label{MM2}
 \end{equation}
and the rotation tensor (7) is
 \begin{equation}
 M_W^{\scriptscriptstyle-}= \pmatrix
  {            0& 14.5_{(5.1)}\cr
   -14.5_{(5.1)}& 0\cr}.
 \label{MM3}
 \end{equation}
Based on (12), we may conclude that the angular velocity of
solid-body rotation of the Cepheid system around the $X$ axis is
$\Omega_W=M^{\scriptscriptstyle-}_{32}=-15\pm5$~km s$^{-1}$
kpc$^{-1}$. This is the minimum (but more reliable) estimate. If
the deformations ($M^{\scriptscriptstyle+}_{23}$) are assumed to
be also related to the effect under consideration, then the
maximum angular velocity of rotation can be estimated as
 $\Omega_W=M^{\scriptscriptstyle-}_{32}-M^{\scriptscriptstyle+}_{23}=-27\pm10$~km
s$^{-1}$ kpc$^{-1}$.

It is important that the direction of the rotation found (minus
sign) is in agreement with the result of our analysis of the
proper motions for red-giant clump stars (Bobylev 2010), where we
used photometric distance estimates with errors
$e_\pi/\pi\approx30\%$. The sign of the angular velocity
$\Omega_W$ depends on the sign of $M_{32}$ (Eqs. (7)--(8)), which
was determined from Cepheids rather reliably owing to the wide
range of coordinates $\Delta Y\approx10$~kpc. Since the range of
coordinates $\Delta Z\approx1.2$~kpc is small when determining
$M_{23}$, the influence of random fluctuations in Cepheid
velocities can be significant.

The value of $\Omega_W=-15\pm5$~km s$^{-1}$ kpc$^{-1}$ derived
from Cepheids exceeds $\Omega_W\approx-4\pm0.5$~km s$^{-1}$
kpc$^{-1}$ obtained from red-giant-clump stars by Bobylev (2010)
by a factor of 4. Such a difference may be related to the sample
ages: the mean age of our sample of Cepheids is 77~Myr, while the
mean age of the red-giant-clump stars is approximately 1~Gyr.
However, this question requires a further study based on larger
volumes of more accurate data.

\section*{CONCLUSIONS}
We considered the space velocities of about 200 long-period (with
periods of more than 5 days) classical Cepheids with known proper
motions and line-of-sight velocities whose distances were
estimated from the period–luminosity relation.

We applied the linear Ogorodnikov–Milne model to analyze their
kinematics. The Galactic rotation that we found based on a more
complex model was excluded from the observed velocities in
advance. Two significant gradients were detected in the Cepheid
velocities:
 $\partial W/\partial Y=-2.1\pm0.7$~km s$^{-1}$ kpc$^{-1}$ and
 $\partial V/\partial Z=  27\pm10 $~km s$^{-1}$ kpc$^{-1}$.
 This leads us to conclude that the angular
velocity of solid-body rotation around the Galactic x axis is
 $\Omega_W=-15\pm5$~km s$^{-1}$ kpc$^{-1}$, which we associate with
a manifestation of the warp of the stellar--gaseous Galactic disk.
Indeed, the relationship between the spatial distribution of
Cepheids and the warp of the stellar--gaseous Galactic disk may be
considered to have been firmly established (Fernie 1968; Berdnikov
1987; Bobylev 2013). The results of our study show that the
kinematic relationship of Cepheids to this phenomenon is also
highly likely.

The method considered here can be useful for a future analysis of
large volumes of data, for example, from the GAIA space experiment
or on masers with their trigonometric parallaxes measured by VLBI.

\subsection*{ACKNOWLEDGMENTS}
We are grateful to the referee for
helpful remarks that contributed to a improvement of the paper.
This work was supported by the ``Nonstationary Phenomena in
Objects of the Universe'' Program of the Presidium of the Russian
Academy of Sciences and the ``Multiwavelength Astrophysical
Research'' grant no. NSh--16245.2012.2 from the President of the
Russian Federation.

\section*{REFERENCES}
{\small

\quad~~1. A.A. Acharova, Yu.N. Mishurov, and V.V. Kovtyukh, Mon.
Not. R. Astron. Soc. 420, 1590 (2012).

2. J. Bailin, Astrophys. J. 583, L79 (2003).

3. E. Battaner, E. Florido, and M.L. Sanchez-Saavedra, Astron.
Astrophys. 236, 1 (1990).

4. L.N. Berdnikov, Astron. Lett. 13, 45 (1987).

5. L.N. Berdnikov, A.K. Dambis, and O.V. Vozyakova, Astron.
Astrophys. Suppl. Ser. 143, 211 (2000).

6. V.V. Bobylev, Astron. Lett. 36, 634 (2010).

7. V.V. Bobylev, Astron. Lett. 39, 95 (2013).

8. V.V. Bobylev and A.T. Bajkova, Astron. Lett. 38, 638 (2012).

9. V.V. Bobylev, A.T. Bajkova, and A.S. Stepanishchev, Astron.
Lett. 34, 515 (2008).

10. J.C. Cersosimo, S. Mader, N.S. Figueroa, et al., Astrophys. J.
699, 469 (2009).

11. R. Drimmel and D.N. Spergel, Astrophys. J. 556, 181 (2001).

12. R. Drimmel, R.L. Smart, and M.G. Lattanzi, Astron. Astrophys.
354, 67 (2000).

13. M. Feast and P.Whitelock, Mon. Not. R. Astron. Soc. 291, 683
(1997).

14. J.D. Fernie, Astron. J. 73, 995 (1968).

15. T. Foster and B. Cooper, ASP Conf. Ser. 438, 16 (2010).

16. P. Fouqu, P. Arriagada, J. Storm, et al., Astron. Astrophys.
476, 73 (2007).

17. G.A. Gontcharov, Astron. Lett. 32, 795 (2006).

18. The Hipparcos and Tycho Catalogues, ESA SP-1200 (1997).

19. E. Hog, C. Fabricius, V.V. Makarov, et al., Astron. Astrophys.
355, L27 (2000).

20. P.M.W. Kalberla and L. Dedes, Astron. Astrophys. 487, 951
(2008).

21. E.V. Kazarovets, N.N. Samus’, O.V. Durlevich, et al., Astron.
Rep. 53, 1013 (2009).

22. M. L\'opez-Corredoira, J. Betancort-Rijo, and J. Beckman,
Astron. Astrophys. 386, 169 (2002).

23. Yu.N. Mishurov, I.A. Zenina, A.K. Dambis, et al., Astron.
Astrophys. 323, 775 (1997).

24. M. Miyamoto and Z. Zhu, Astron. J. 115, 1483 (1998).

25. M. Miyamoto, M. S\^oma, and M. Yoshizawa, Astron. J. 105, 2138
(1993).

26. Y. Momany, S. Zaggia, G. Gilmore, et al., Astron. Astrophys.
451, 515 (2006).

27. K.F. Ogorodnikov, Dynamics of Stellar Systems (Fizmatgiz,
Moscow, 1965) [in Russian].

28. C.A. Olano, Astron. Astrophys. 423, 895 (2004).

29. L. Sparke and S. Casertano, Mon. Not. R. Astron. Soc. 234, 873
(1988).

30. G. Westerhout, Bull. Astron. Inst. Netherlands 13, 201 (1957).

31. I. Yusifov, astro-ph/0405517 (2004).

32. N. Zacharias, C. Finch, T. Girard, et al., Astron. J. 145, 44
(2013).

\end{document}